\documentstyle[seceq,preprint]{ptptex}

\newcommand{\nn}{\nonumber\\}

\newcommand{\ket}[1]{\left| #1 \right>}
\newcommand{\abs}[1]{\left| #1 \right|}
\newcommand{\rL}{{\rm L}}
\newcommand{\rR}{{\rm R}}
\newcommand{\Q}{Q_{\rm B}}
\newcommand{\half}{\frac{1}{2}}
\newcommand{\JHEP}{J.~High~Energy~Phys.\ }
\preprintnumber[3cm]{
hep-th/0107046}

\title{
Wilson Lines and Classical Solutions\\ in Cubic Open String Field Theory
}

\author{
Tomohiko {\sc Takahashi}\footnote{E-mail: tomo@asuka.phys.nara-wu.ac.jp} 
and Seriko {\sc Tanimoto}\footnote{E-mail: tanimoto@asuka.phys.nara-wu.ac.jp}
}

\inst{
Department of Physics, Nara Women's University, Nara 630-8506, Japan
}

\recdate{
July 6, 2001
}

\abst{
We construct exact classical solutions in cubic open string field
theory. Through the redefinition of the string field, we find that
the solutions correspond to finite deformations of the Wilson lines.
The solutions have well-defined Fock space expressions,
and they have no branch cut singularity of marginal parameters,
which was found in the analysis using a level truncation approximation in
the Feynman-Siegel gauge. We also discuss marginal tachyon lump
solutions at critical radius.
}

\begin{document}

\maketitle

\section{Introduction}

Cubic open string field theory\cite{rf:CSFT} has been 
a useful tool to understand 
various conjectures
about tachyon condensation in string
theory.\cite{rf:Sen1,rf:Sen2,rf:Sen3}
Most works on this subject
have been carried
out using a level truncation scheme in Feynman-Siegel
gauge.\cite{rf:SZ-tachyon,rf:KS,rf:MT} 
It has been shown that the tachyon vacuum solution in the level
truncation scheme 
is BRS-invariant and that it satisfies the equation of motion outside
the Feynman-Siegel gauge.\cite{rf:HS}
The validity of the Feynman-Siegel gauge has been checked also
for the tachyon lump solution.\cite{rf:MS}
However, we are unable to understand global
structure of the vacuum in the string field theory
because of branch cut singularities due to the Feynman-Siegel
gauge.\cite{rf:SZ-tachyon,rf:KS,rf:MT}
To gain further
insight into the vacuum structure, it is necessary to construct
exact solutions without fixing 
the Feynman-Siegel gauge.

In order to obtain exact results in cubic open string field theory,
we consider 
finite deformations of the Wilson
lines background, as situation somewhat simpler than that of the 
tachyon vacuum. The classical solutions corresponding to
marginal condensation have been constructed formally in string
field theories with joining-splitting type
vertices.\cite{rf:KZ,rf:Yoneya,rf:HN,rf:Kawano,rf:KT2}
We expect that this previously obtained knowledge will be helpful in
constructing exact solutions 
of the condensation of marginal operators
in cubic open string field theory.

There is a problem involved in the study of the Wilson lines
deformations
in cubic open string field theory.
According to the calculation employing the level truncation
scheme,\cite{rf:SZ-LM}
it is impossible to carry out an analysis outside a
critical value of marginal parameters,
though, to some extent, we can describe string field configurations
representing the 
condensations 
of gauge fields.
Interestingly, however, from level truncated analysis
for tachyon condensation,
it has been shown that
singularities in the tachyon potential originate in the
Feynman-Siegel gauge.\cite{rf:ET}
This gives us reason to believe that we may be able to
obtain solutions beyond the critical value if we choose
something other than the Feynman-Siegel gauge. It also seems possible
that we might need
different coordinate patches to describe the full moduli space with the
string field theory.\cite{rf:SZ-LM}

In this paper, we construct exact solutions in cubic open string
field theory, without fixing any gauge. We show that the solutions
correspond to 
deformations of the Wilson lines background through a homogeneous
redefinition 
of the string field. 
We find that there are parameters in the solutions corresponding to
the vacuum expectation value of the gauge field, but the solutions have
no singularity of the parameters. We conclude that the
singularity in the level truncation scheme is a gauge artifact.

We also construct solutions of a
marginal tachyon lump.\cite{rf:Sen3,rf:SZ-LM}
We find that at the critical compactified radius, a tachyon lump
is generated by marginal deformations of $U(1)$ currents,
and it corresponds to
the Wilson line operator.\cite{rf:Sen3}
In contrast to the situation for Wilson line deformations,
we are able to obtain tachyon lump solutions.

The rest of this paper is organized as follows. In \S 2 we
construct the exact solutions and redefine the string field.
Then, we evaluate a solution using the oscillator expression to find that
the solutions have well-defined Fock space expressions. 
In \S 3, we discuss the solutions corresponding to the marginal
tachyon lump. We give a summary in \S 4.
In the Appendix, we prove formulas related to the delta function.

\section{Exact solutions in cubic open string field theory}

For simplicity, we single out a direction that is tangential to
D-branes. 
The canonical commutation relation of the string coordinate
and momentum is given by
\begin{eqnarray}
\label{Eq:cancom}
\left[X(\sigma),\,P(\sigma')\right]=i\delta(\sigma,\sigma'),
\end{eqnarray}
and others are zero.
The delta function is defined by
\begin{eqnarray}
\label{Eq:delta}
\delta(\sigma,\sigma')=\frac{1}{\pi}+\frac{2}{\pi}
\sum_{n\geq 1}\cos(\sigma)\cos(\sigma').
\end{eqnarray}
We expand $X(\sigma)$ and $P(\sigma)$ in the oscillator modes as
\begin{eqnarray}
\label{Eq:modexpXP}
X(\sigma)=x+i\sqrt{2\alpha'}\sum_{n\neq 0}\frac{1}{n}
\alpha_n\cos(n\sigma),\nn
P(\sigma)=\frac{p}{\pi}+\frac{1}{\sqrt{2\alpha'}\pi}
\sum_{n\neq 0}\alpha_n\cos(n\sigma).
\end{eqnarray}
The mode expansions of ghost variables are given by
\begin{eqnarray}
\label{Eq:modexpCB}
c(\sigma)=\sum_n c_n e^{-in\sigma},\ \ \ 
b(\sigma)=\sum_n b_n e^{-in\sigma}.
\end{eqnarray}

In order to construct solutions in the string field theory, we define
the operators
\begin{eqnarray}
&&V_\rL(f)=\frac{2}{\pi}\int_0^{\frac{\pi}{2}} 
   d\sigma f(\sigma) V(\sigma),\ \ \ 
V_\rR(f)=\frac{2}{\pi}\int_{\frac{\pi}{2}}^\pi 
   d\sigma f(\sigma) V(\sigma),\nn
&&C_\rL(f)=\frac{2}{\pi}\int_0^{\frac{\pi}{2}} 
   d\sigma f(\sigma) C(\sigma),\ \ \ 
C_\rR(f)=\frac{2}{\pi}\int_{\frac{\pi}{2}}^\pi
   d\sigma f(\sigma) C(\sigma),
\end{eqnarray}
where $f(\sigma)$ is an arbitrary function. The quantities $V(\sigma)$
and $C(\sigma)$ are 
defined by
\begin{eqnarray}
\label{Eq:VCdef}
&&V(\sigma)=\frac{1}{2\sqrt{2\alpha'}}
\left[c(\sigma)\left(2\pi\alpha' P(\sigma)+X'(\sigma)\right)
+c(-\sigma)\left(2\pi\alpha' P(\sigma)-X'(\sigma)\right)
\right],\nn
&&C(\sigma)=\frac{1}{2}\left(c(\sigma)+c(-\sigma)\right).
\end{eqnarray}
The operator $V$ corresponds to the vector vertex operator with zero
momentum. Substituting Eqs.~(\ref{Eq:modexpXP}) and (\ref{Eq:modexpCB})
into Eq.~(\ref{Eq:VCdef}), the oscillator expressions of the operators
$V$ and $C$ are found to be
\begin{eqnarray}
V(\sigma)=\sum_m\left(\sum_n c_{m-n} \alpha_n\right) \cos(m\sigma),\ \ \ 
C(\sigma)=\sum_n c_n \cos(n\sigma).
\end{eqnarray}
The operator as $V$ can be written in terms of the commutator of the BRS charge
and the string coordinate
\begin{eqnarray}
\label{Eq:QV}
V(\sigma)=\frac{i}{\sqrt{2\alpha'}}\left[\Q,\,X(\sigma)\right].
\end{eqnarray}

We can find that the commutation relations of $V_{\rL(\rR)}$ and
$C_{\rL(\rR)}$ are given by
\begin{eqnarray}
\label{Eq:VLVL}
&&\left\{V_\rL(f),\,V_\rL(g)\right\}=-2\left\{\Q,\,C_\rL(fg)\right\},\nn
&&\left\{V_\rR(f),\,V_\rR(g)\right\}=-2\left\{\Q,\,C_\rR(fg)\right\},
\end{eqnarray}
while other commutation relations vanish. Here, 
Neumann boundary conditions are imposed on the functions
$f(\sigma)$ and $g(\sigma)$. 
Using the formulas involving the delta
function (see in Appendix A)
\begin{eqnarray}
\label{Eq:deltaformula}
&&
\int_0^{\frac{\pi}{2}}d\sigma\int_0^{\frac{\pi}{2}}d\sigma'
f(\sigma)g(\sigma')\delta(\sigma,\sigma')
=\int_0^{\frac{\pi}{2}}d\sigma
f(\sigma) g(\sigma), \nn
&&
\int_{\frac{\pi}{2}}^\pi d\sigma\int_{\frac{\pi}{2}}^\pi d\sigma'
f(\sigma)g(\sigma')\delta(\sigma,\sigma')
=\int_{\frac{\pi}{2}}^\pi d\sigma
f(\sigma) g(\sigma), \nn
&&
\int_0^{\frac{\pi}{2}}d\sigma\int_{\frac{\pi}{2}}^\pi d\sigma'
f(\sigma)g(\sigma')\delta(\sigma,\sigma')
=0,
\end{eqnarray}
we can calculate the above commutation relations.
For example, we have
\begin{eqnarray}
\label{Eq:precom}
\left\{V_\rL(f),\,V_\rL(g)\right\}
&=& \frac{2 i}{\sqrt{2\alpha'}\pi}\int^{\frac{\pi}{2}}_0 d\sigma' g(\sigma')
\left\{V_\rL(f),\,\left[\Q,\,X(\sigma')\right]\right\}\nn
&=& -\frac{4}{\pi}
 \left\{\Q, \int_0^{\frac{\pi}{2}}d\sigma\int_0^{\frac{\pi}{2}}d\sigma'
f(\sigma)g(\sigma')C(\sigma)\delta(\sigma,\sigma')\right\} \nn
&=&
-\frac{4}{\pi}
\left\{\Q, \int_0^{\frac{\pi}{2}}d\sigma
f(\sigma)g(\sigma)C(\sigma)\right\} \nn
&=&
-2\left\{\Q, C_\rL(fg)\right\}. 
\end{eqnarray}

From the connection conditions on the identity
string field and the vertex,
we find the following equations.
For any pair of string fields $A$ and $B$,
\begin{eqnarray}
&&\label{Eq:VI}
V_\rL(h) I= V_\rR(h) I,\\
&&\label{Eq:VAB}
(V_\rR(h)A)*B=(-)^{\abs{A}}\,A*(V_\rL(h) B),\\
&&\label{Eq:CI}
C_\rL(h) I= -C_\rR(h) I,\\
&&\label{Eq:CAB}
(C_\rR(h)A)*B=-(-)^{\abs{A}}\,A*(C_\rL(h) B),
\end{eqnarray}
where $I$ denotes the identity string field
and $h(\sigma)$ satisfies $h(\pi-\sigma)=h(\sigma)$.
In Eq.~(\ref{Eq:CI}), $h(\sigma)$ is more strongly constrained, so that
$C_{\rL(\rR)}$ is a regular operator on the identity string field,
because the ghost operator has a singularity at the midpoint. We can
find such 
a function $h$ using the 
oscillator expression of the identity string
field given below,
though it can also be obtained with a more elegant procedure.\cite{rf:RZ}

Now, we can show that exact solutions in the cubic open string field
theory are given by
\begin{eqnarray}
\label{Eq:solution}
 \Psi_0(\lambda) = \sqrt{2\alpha'} a_i V_\rL(\lambda) I
+2 \alpha' {a_i}^2 C_\rL(\lambda^2) I,
\end{eqnarray}
where the $a_i$ are real parameters, where $i$ is the Chan-Paton index,
\footnote{We omit $\delta_{ij}$ in the identity string field.}
and the function $\lambda$ satisfies 
Neumann boundary conditions and
$\lambda(\pi-\sigma)=\lambda(\sigma)$, explicitly
$\lambda(\sigma)=\sum_n \lambda_n \cos(2n\sigma)$.
Furthermore, the function $\lambda(\sigma)$ must be restricted to
make $C_{\rL}(\lambda^2)$ a well-defined operator at the
midpoint of the 
identity string field.
From Eq.~(\ref{Eq:QV}), we find
\begin{eqnarray}
 \Q \Psi_0 = 2\alpha' {a_i}^2 \Q C_\rL(\lambda^2) I.
\end{eqnarray}
Using the properties of $V_{\rL(\rR)}$ and $C_{\rL(\rR)}$,
Eqs.~(\ref{Eq:VLVL}) and (\ref{Eq:VI})--(\ref{Eq:CAB}), we find 
\begin{eqnarray}
 \Psi_0*\Psi_0 = 2\alpha' {a_i}^2 V_\rL(\lambda)^2 I
 = -2\alpha' {a_i}^2 \Q C_\rL(\lambda^2) I.
\end{eqnarray}
As a result, $\Psi_0$ obeys the equation of motion,
$\Q\Psi_0+\Psi_0*\Psi_0=0$.

Since $\Psi_0$ is a solution for any values of the $a_i$, the potential
energy $V(\Psi_0)=-S(\Psi_0)$ is equal to that for $a_i=0$. Therefore,
the potential energy is alway 0.\cite{rf:KZ}
This implies that the parameters $a_i$ correspond to marginal parameters.

If we expand the string field around the solution as
$\Psi=\Psi_0+\Psi'$,
the BRS charge of the resultant theory is given by
$\Q'= \Q+D_{\Psi_0}$,
where $D_{\Psi_0}$ is defined by $D_{\Psi_0} A = \Psi_0*A-(-)^{\abs{A}}
A*\Psi_0$ for any string field $A$. From the properties of $V_{\rL(\rR)}$
and $C_{\rL(\rR)}$, we find that the BRS charge in the shifted theory
is given by
\begin{eqnarray}
\hspace{-1cm}
\Q'=\Q+\sqrt{2\alpha'}\left(a_i
V_\rL(\lambda)-a_j V_\rR(\lambda)\right)
+2 \alpha'\left({a_i}^2 C_\rL(\lambda^2)
+{a_j}^2 C_\rR(\lambda^2)\right).  
\end{eqnarray}
We can easily check the nilpotency of the shifted BRS charge
by using Eq.~(\ref{Eq:VLVL}). 

Let us consider the redefinition of the shifted string field. To do so,
we introduce  
the operators
\begin{eqnarray}
 X_\rL(f)=\frac{2}{\pi}\int_0^{\frac{\pi}{2}} d\sigma
   f(\sigma) X(\sigma),\ \ \ 
 X_\rR(f)=\frac{2}{\pi}\int_{\frac{\pi}{2}}^\pi d\sigma
   f(\sigma) X(\sigma).
\end{eqnarray}
We find the commutators
\begin{eqnarray}
\label{Eq:XQV}
\left[X_\rL(f),\, \Q\right]=i\sqrt{2\alpha'} V_\rL(f),
\ \ \ \left[X_\rL(f),\, V_\rL(g)\right]=i2\sqrt{2\alpha'} C_\rL(fg).
\end{eqnarray}
From Eq.~(\ref{Eq:XQV}), it follows that the shifted BRS charge
can be transformed into the original one as
\begin{eqnarray}
 e^{iB(\lambda)} \Q' e^{-iB(\lambda)} =\Q,
\end{eqnarray}
where the operator $B$ is defined by
\begin{eqnarray}
 B(f) = a_i X_\rL(f)-a_j X_\rR(f).
\end{eqnarray}
By the connection conditions of the string coordinate
on the vertex, the cubic term of the action is invariant
under the transformation of the string field generated by the operator
$B$
\begin{eqnarray}
 \int (e^{-iB(\lambda)} \Phi)*(e^{-iB(\lambda)} \Psi)
*(e^{-iB(\lambda)}\Xi)
=\int \Phi*\Psi*\Xi.
\end{eqnarray}
The invariance is generalized to the $n$-string vertices
of the midpoint interaction.

Consequently, if we redefine the string field as
$\tilde{\Psi}=e^{iB}\Psi'$, 
the form of the shifted action apparently becomes the original
one. However, since if $\lambda(\sigma)$ has a zero mode,  
the operator $e^{iB}$ contains $\sim\hspace{-.3em}e^{i(a_i-a_j) x}$ as
the zero mode part, 
the component fields in the redefined string field are no
longer periodic if the relevant direction is compactified. The effect of
the $a_i$ on the periodicity implies that the redefined theory represents
the strings in 
the Wilson lines 
background. Hence, the parameters $a_i$ correspond to
the finite vacuum expectation values of the gauge field.
 
we now consider the oscillator expansions of the solution
and show that the solutions have well-defined Fock space expressions.
The operator expression of the identity string field
is given by\cite{rf:GJ-1,rf:GJ-2}
\begin{eqnarray}
&&
\ket{I}=\frac{1}{4i}b\left(\frac{\pi}{2}\right)
b\left(-\frac{\pi}{2}\right)
e^{E'}c_0 c_1\ket{0},\nn
&&
E'=\sum_{n\geq 1}(-)^n\left[-\frac{1}{2n}\alpha_{-n}\cdot\alpha_{-n}
+c_{-n}b_{-n}\right],
\end{eqnarray}
where $\ket{0}$ is $SL(2,{\rm R})$ invariant vacuum. For convenience, we
rewrite 
it \footnote{Expanding this expression, we can check that the first
few levels of terms agree with the expression obtained using the
Virasoro generators in Ref.~\citen{rf:RZ}.
}
\begin{eqnarray}
\label{Eq:I}
&&
\ket{I}=e^E\ket{0}, \nn
&&
E=-\sum_{n\geq 1}\frac{(-)^n}{2n}\alpha_{-n}\cdot\alpha_{-n}
+\sum_{n\geq 2}(-)^n c_{-n}b_{-n} \nn
&&\ \ \ \ \ 
-2 c_0 \sum_{n\geq 1}(-)^n b_{-2n}-(c_1-c_{-1})
\sum_{n\geq 1}(-)^n b_{-2n-1}.
\end{eqnarray}
From Eq.~(\ref{Eq:I}), we find directly
\begin{eqnarray}
&& \left(c_0-\frac{(-)^n}{2}(c_{2n}+c_{-2n})\right) \ket{I}=0, \nn
&& \left(c_1-c_{-1}-(-)^n(c_{2n+1}-c_{-2n-1})\right) \ket{I}=0,
\end{eqnarray}
where $n=1,2,\cdots$. We can easily check that the first few equations
agree with the result of Ref.~\citen{rf:RZ}. 
Applying the ghost operator to the expression in Eq.~(\ref{Eq:I}),
we can evaluate the midpoint singularity on the identity string field.
We obtain
\begin{eqnarray}
c(\sigma)\ket{I}=&&\left[i\,c_0\,\tan \sigma+c_1\,\frac{1}{2\cos\sigma} 
+c_{-1}\,\frac{1+2\cos(2\sigma)}{\cos\sigma}  \right. \nn
&&+2 \sum_{n\geq 1}\left(i\,c_{-2n}\,\sin(2n\sigma) 
+c_{-2n-1}\,\cos[(2n+1)\sigma] \right)
\Big]\ket{I}.
\end{eqnarray}
As the simplest
example, we consider a solution for $\lambda(\sigma)=1+\cos(2\sigma)$.
In this case, we can expand the solution up to level two as follows,
\begin{eqnarray}
 \ket{\Psi_0} =&& \left[\sqrt{2\alpha'}a_i \left(
 c_1\alpha_{-1}+2c_{-1}\alpha_{-1}+\half c_1 \,\alpha_{-1}
(\alpha_{-1}\cdot\alpha_{-1})\right) 
\right.\nn
&&\left.
\ \ \ +2\alpha' {a_i^2}
\left(\frac{8}{3\pi}c_1+\frac{88}{15\pi}c_{-1}
+\frac{4}{3\pi}c_1(\alpha_{-1}\cdot\alpha_{-1})\right)\right]\ket{0} \nn
&&
\hspace{-1cm}
+c_0\left[
-\sqrt{2\alpha'}\,a_i \left(\alpha_{-2}  
+2c_1 b_{-2}\alpha_{-1} \right)
- 2\alpha' {a_i}^2 \,\frac{16}{3\pi} c_1 b_{-2}\right]\ket{0}+\cdots.
\end{eqnarray}
It turns out that the solution of Eq.~(\ref{Eq:solution}) is
well-defined in the level truncation scheme, and it is outside the
Feynman-Siegel gauge because of
the term proportional to $c_0$.

\section{Marginal tachyon lump solutions}

If the compactified radius has the critical value, $R=\sqrt{\alpha'}$,
a tachyon lump is generated by a marginal
operator.\cite{rf:Sen3,rf:SZ-tachyon}
Let us consider the construction of tachyon lump solutions at the
critical radius. 
In the solutions of Eq.~(\ref{Eq:solution}), $V_\rL$ is constructed
by the marginal operator $\partial X$ as $V_\rL\sim \int c \partial X$.
Therefore, we have only to replace $\partial X$ with a $U(1)$ current in order
to obtain a solution corresponding to a marginal deformation of the $U(1)$
current. 

Here we give tachyon lump solutions. We give the mode
expansion of the marginal operator
corresponding to  tachyon lumps as
\begin{eqnarray}
t(z)
=\sqrt{2}\cos\left(\frac{X(z)}{\sqrt{\alpha'}}\right)
=\sum_n t_n z^{-n-1},
\end{eqnarray}
where $X(z)$ is defined by
\begin{eqnarray}
X(z)=x-2i \alpha'p \ln z
+i\sqrt{2\alpha'}\sum_{n\neq 0}\frac{1}{n}\alpha_n z^{-n}.
\end{eqnarray}
Then, we find that $t_n$ satisfies the same commutation relations as
$\alpha_n$:
\begin{eqnarray}
\label{Eq:tcom}
&&[t_m,\,t_n]=m\delta_{m+n}.
\end{eqnarray}
Since the operator $t(z)$ is a $U(1)$ current, we can transform $t(z)$
into an operator in the $\rho$ plane by the mapping
$\rho=\tau+i\sigma=\ln z$. 
Using the operator at $\tau=0$
\begin{eqnarray}
 t(\sigma)=\sum_n t_n e^{-in\sigma},
\end{eqnarray}
we define operators similar to $V_{\rL(\rR)}$ by
\begin{eqnarray}
 T_\rL(f)= 
  \frac{2}{\pi}\int_0^{\frac{\pi}{2}} d\sigma f(\sigma)T(\sigma),\ \ \ 
 T_\rR(f)= 
  \frac{2}{\pi}\int_{\frac{\pi}{2}}^\pi d\sigma f(\sigma)T(\sigma),
\end{eqnarray}
where $T(\sigma)$ is given by
\begin{eqnarray}
 T(\sigma)&=&\half\left(c(\sigma)t(\sigma)+c(-\sigma)t(-\sigma)\right)\nn
 &=& \sum_m\left(\sum_n c_{m-n}t_n\right)\cos(m\sigma).
\end{eqnarray}
Since $t_m$ satisfies the same commutation relations as $\alpha_m$ given
in
Eq.~(\ref{Eq:tcom}), algebras similar to those of
$V_{\rL(\rR)}$ and $C_{\rL(\rR)}$ follow. Indeed, we have
\begin{eqnarray}
\label{Eq:TLTL}
&&\left\{T_\rL(f),\,T_\rL(g)\right\}=-2\left\{\Q,\,C_\rL(fg)\right\},\nn
&&\left\{T_\rR(f),\,T_\rR(g)\right\}=-2\left\{\Q,\,C_\rR(fg)\right\}.
\end{eqnarray}
Since $t(\sigma)$ is a $U(1)$ current, it satisfies the same
connection conditions on the vertex
and on the identity string field as $P(\sigma)$ and $X'(\sigma)$.
Therefore, we also obtain the similar identities related to $I$ and  
the $*$ product:
\begin{eqnarray}
 &&\label{Eq:TI}
T_\rL(h) I= T_\rR(h) I,\\
&&\label{Eq:TAB}
(T_\rR(h)A)*B=(-)^{\abs{A}}\,A*(T_\rL(h) B).
\end{eqnarray}

From Eqs.~(\ref{Eq:TLTL}), (\ref{Eq:TI}) and (\ref{Eq:TAB}), 
we find that the solutions corresponding to marginal tachyon lumps
are given by
\begin{eqnarray}
 T_0(\lambda) = s_i\,T_\rL(\lambda) I + {s_i}^2\,C_\rL(\lambda^2) I,
\end{eqnarray}
where the $s_i$ correspond to marginal parameters.
If we choose $\lambda(\sigma)=1+\cos(2\sigma)$, the expanded form of the
solution 
up to level two is given by
\begin{eqnarray}
 \ket{T_0} =&& \left[s_i \left(
 c_1t_{-1}+2c_{-1}t_{-1}+\half c_1 \,t_{-1}(\alpha_{-1}\cdot\alpha_{-1})\right)
\right.\nn
&&\left.
\ \ \ +{s_i}^2
\left(\frac{8}{3\pi}c_1+\frac{88}{15\pi}c_{-1}
+\frac{4}{3\pi}c_1(\alpha_{-1}\cdot\alpha_{-1})\right)\right]\ket{0} \nn
&&
+c_0\left[
-s_i\, \left(t_{-2} + 2c_1 b_{-2}t_{-1}\right)
- {s_i}^2 \,\frac{16}{3\pi} c_1 b_{-2}\right]\ket{0}+\cdots.
\end{eqnarray}

\section{Summary}

We constructed exact solutions in cubic open string field theory
without choosing the Feynman-Siegel gauge. There are many solutions
associated with the functions $\lambda(\sigma)$ that make the operator
$C_{\rL(\rR)}$ well-defined on the identity string field.
We showed that through the redefinition of the string field,
the shifted string field theory becomes a theory with finite
Wilson lines deformations for any function
$\lambda(\sigma)$ with a zero mode.  
The solutions we constructed have no branch cut singularity. The branch
singularity 
previously found is revealed to be a gauge artifact.
Our solutions have well-defined Fock space expressions.  

It is difficult to construct a solution of the
tachyon vacuum and to redefine a string field to obtain a theory
without physical open string excitations, which is expected 
to be the case for
the vacuum string
field theory.\cite{rf:VSFT-1,rf:VSFT-2,rf:VSFT-3,rf:VSFT-4,rf:VSFT-5}
Though we can solve the equation of motion exactly for the Wilson lines
condensation, it seems that we are unable to apply the method
of the marginal case directly to the tachyon condensation.

Our solutions are well-defined in the Fock space.
The finite marginal solutions in the light-cone type
string field theories\cite{rf:KZ} are rather formal, and it is very
difficult
to represent them in terms of the oscillator expressions. However, the
solution of 
the dilaton condensation provides the correct result for the shift
of the string coupling constant in light-cone type string field
theories.\cite{rf:Yoneya,rf:HN,rf:Kawano} We have been still faced with the
problem of how to treat finite solutions in the light-cone type string
field theories.\cite{rf:KZ}
\vspace{1cm}


\section*{Acknowledgements}
e would like to thank H.~Hata, T.~Kugo, B.~Zwiebach and T.~Matsuyama
for useful comments. 
We are also grateful to our colleagues at Nara Women's University
for encouragement.

\appendix
\section{Proof of Eq.~(\ref{Eq:deltaformula})}

We consider the integral
\begin{eqnarray}
 I_{mn} &=& \int_0^{\frac{\pi}{2}}d\sigma 
\int_0^{\frac{\pi}{2}}d\sigma' \cos(m\sigma)
\cos(n\sigma')\delta(\sigma,\sigma')\nn
&=& \frac{1}{\pi}\sum_l
\int_0^{\frac{\pi}{2}}d\sigma \cos(m\sigma)\cos(l\sigma)
\int_0^{\frac{\pi}{2}}d\sigma' \cos(n\sigma')\cos(l\sigma').
\end{eqnarray}
In the case $m\pm n\neq  0$, this becomes
\begin{eqnarray}
 I_{mn}=&&\half\left[\frac{1}{m+n}\sin\left(\frac{(m+n)\pi}{2}\right)
+\frac{1}{m-n}\sin\left(\frac{(m-n)\pi}{2}\right)\right]\nn
&& +\frac{1}{2\pi}\sum_{l\neq m\pm n}
\left[\frac{1}{l(l-m+n)}\sin\left(\frac{l\pi}{2}\right)
                       \sin\left(\frac{(l-m+n)\pi}{2}\right)\right.\nn
&&
\left.
+\frac{1}{l(l-m-n)}\sin\left(\frac{l\pi}{2}\right)
                       \sin\left(\frac{(l-m-n)\pi}{2}\right)\right].
\end{eqnarray}
Summing the above series by using 
\begin{eqnarray}
 \sum_{n\neq 0,m}\frac{1}{n(n-m)}
\sin\left(\frac{n\pi}{2}\right)\sin\left(\frac{(n-m)\pi}{2}\right)
=\frac{\pi^2}{4}\delta_{m,0},
\end{eqnarray}
we evaluate the integral as
\begin{eqnarray}
\hspace{-4em} I_{mn}=\frac{1}{2(m+n)}\sin\left(\frac{(m+n)\pi}{2}\right)
+\frac{1}{2(m-n)}\sin\left(\frac{(m-n)\pi}{2}\right),\ 
(m\pm n\neq 0).
\end{eqnarray}
If $m=n=0$ or $m=\pm n\neq 0$, we find, after similar calculations,
\begin{eqnarray}
 I_{00} &=& \frac{\pi}{2},\nn
I_{nn} &=& I_{n,-n} = \frac{\pi}{4}.\ \ \ 
(n\neq 0).
\end{eqnarray} 

If $f(\sigma)$ and $g(\sigma)$ satisfy Neumann boundary conditions,
we can expand the functions as $f(\sigma)=\sum_n f_n \cos(n\sigma)$
and $g(\sigma)=\sum_n g_n \cos(n\sigma)$.
Then, 
we can easily prove one of the expressions in Eq.~(\ref{Eq:deltaformula}):
\begin{eqnarray}
\label{Eq:del-1}
\int_0^{\frac{\pi}{2}}d\sigma 
\int_0^{\frac{\pi}{2}}d\sigma'f(\sigma) g(\sigma')\delta(\sigma,\sigma')
&=& \sum_{mn}f_m g_n I_{mn}
=\int_0^{\frac{\pi}{2}}d\sigma f(\sigma)g(\sigma).
\end{eqnarray}
By the definition of the delta function, it follows that
\begin{eqnarray}
\label{Eq:del-2}
 \int_0^{\frac{\pi}{2}}d\sigma 
\int_0^\pi d\sigma' f(\sigma)g(\sigma')\delta(\sigma,\sigma')
=\int_0^{\frac{\pi}{2}}d\sigma f(\sigma)g(\sigma).
\end{eqnarray}
Considering Eqs.~(\ref{Eq:del-1}) and (\ref{Eq:del-2}), we can obtain
the other two expressions of Eq.~(\ref{Eq:deltaformula}).



\vspace{1cm}

\noindent
\footnotesize
{\bf Note added:}\footnote{This information was provided to us by H.~Hata
in private communication.}

The solutions of Eq.~(\ref{Eq:solution}) are locally pure gauge.
If we choose as a gauge parameter functional
\begin{eqnarray}
 \Lambda=-i a_i X_\rL(\lambda) I, \nonumber
\end{eqnarray}
the solutions are generated by the gauge transformation of the
zero string field
\begin{eqnarray}
 e^\Lambda *\Q e^{-\Lambda} &=& 
e^{-i a_i X_\rL(\lambda)}I*\Q e^{i a_i X_\rL(\lambda)} I \nn
&=& e^{-i a_i X_\rL(\lambda)}\Q e^{i a_i X_\rL(\lambda)} I \nn
&=&  \sqrt{2\alpha'} a_i V_\rL(\lambda) I
+2 \alpha' {a_i}^2 C_\rL(\lambda^2) I. \nonumber
\end{eqnarray}

\end{document}